\documentclass[twoside,a4paper,11pt]{proceedings}
\usepackage{graphicx}
\usepackage{hyperref}
\usepackage{movie15}
\usepackage{natbib}
\usepackage{caption}
\usepackage{amstext}
\begin{document}
\pagenumbering{arabic}
\pagestyle{myheadings}
\thispagestyle{empty}
\vspace*{-1cm}
{\flushleft\includegraphics[width=3cm,viewport=0 -30 200 -20]{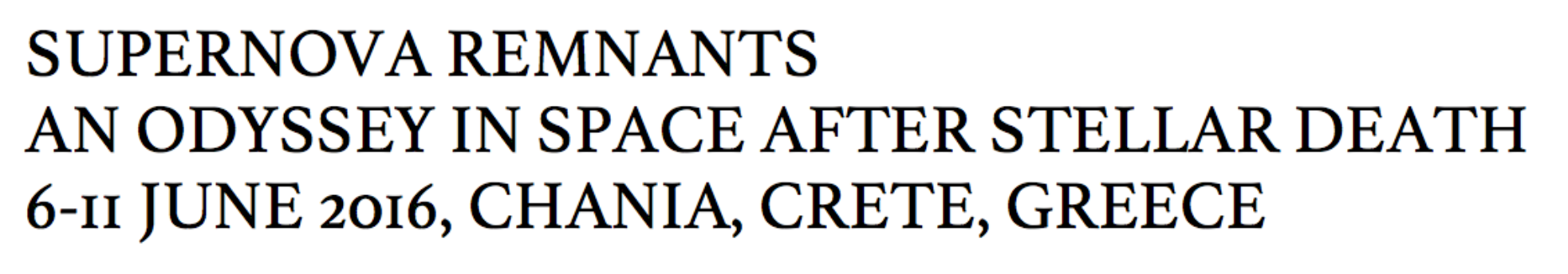}}
\vspace*{0.2cm}
\begin{flushleft}
{\bf {\LARGE
Cygnus Loop: A double bubble? 
}\\
\vspace*{1cm}
J. West$^{1, 2}$,
S. Safi-Harb$^1$,
I. Reichardt$^3$,
J. Stil$^4$,
R. Kothes$^5$,
T. Jaffe$^6$,
and GALFACTS team.
%
}\\
\vspace*{0.5cm}
%
$^{1}$
Dept. of Physics and Astronomy, University of Manitoba, Winnipeg, Canada \\
$^{2}$
Dunlap Institute, University of Toronto, Toronto, Canada  \\
$^{3}$
University of Padova, Italy\\
$^{4}$
Dept. of Physics and Astronomy, University of Calgary, Calgary, Canada\\
$^{5}$
National Research Council of Canada, Penticton, Canada\\
$^{6}$
University of Maryland, College Park, MD, USA, and NASA Goddard Space Flight Center, Greenbelt, MD, USA
%
\end{flushleft}
\markboth{
Cygnus Loop: A double bubble? 
}{
West et al.
}
\thispagestyle{empty}
\vspace*{0.4cm}
\begin{minipage}[l]{0.09\textwidth}
\ 
\end{minipage}
\begin{minipage}[r]{0.9\textwidth}
\vspace{1cm}
\section*{Abstract}{\small
The Cygnus Loop is a well-studied supernova remnant (SNR) that has been observed across the electromagnetic spectrum. Although widely believed to be an SNR shell with a blow-out region in the south, we consider the possibility that this object is two SNRs projected along the same line-of-sight by using multi-wavelength images and modelling. Our results show that a model of two objects including some overlap region/interaction between the two objects has the best match to the observed data.
\vspace{10mm}
\normalsize}
\end{minipage}

\section{Introduction}
The Cygnus Loop is a well-studied, large, bright and nearby supernova remnant (SNR) that has been observed across the electromagnetic spectrum. It is believed to be an SNR shell with a blow-out region in the south. However, it has also been suggested that this object is two SNRs. We consider this two-SNR scenario by using a multi-wavelength view, focusing on new multi-frequency radio polarization data from GALFACTS (the Galactic Arecibo L-band Feed Array Continuum Transit Survey), with the addition of microwave (Planck), infrared (WISE), ultraviolet (GALEX), X-ray (ROSAT), and gamma-ray (Fermi-LAT) data. In addition, we present modelling efforts that support the 2-SNR interpretation.

\section{Analysis}
{\bf Data:} Based on Effelsberg 100-m telescope data at 2.7 GHz, \citet{2002A&A...389L..61U} first suggested that the Cygnus Loop may in fact be two SNRs, observed in projection. We present new radio data from GALFACTS,
a sensitive, high-resolution, spectro-polarimetric survey of the Arecibo sky, along with data at other wavelengths (from radio to gamma-rays) that support this 2-SNR picture. 
The colour multi-wavelength image (Figure~1) clearly shows a distinct difference between the two regions (defined in Table~1 and Figure~2).

The GALFACTS polarized intensity data (Figure~3) shows very distinct polarization properties between the northern and southern regions. The southern region is much more highly polarized when compared to the northern region, which is nearly completely depolarized.

Higher frequency (30~GHz) polarization data from Planck reveals that both regions are polarized, as would be expected from synchrotron radiation. The depolarization at lower frequencies may be due to a Faraday screen existing between the two SNRs (which could depolarize if the northern SNR is behind the screen) and/or a difference in age between the two SNRs, if the northern one is older and more radiative than the southern one.
A spectral analysis of Fermi/LAT data \citep[see also Figure 4]{2015arXiv150203053R} further supports this 2-SNR scenario.

{\bf Model:} We model the 2-SNR scenario by using the coordinate transformation technique applied in \citet{2016A&A...587A.148W}. 
We assume that the northern SNR is at a fixed distance of 500 pc (radius=28 pc), and we vary the distances to the southern SNR, as well as the transition distance for the two magnetic field orientations.
If we interpret the data as showing two superimposed bilateral SNRs, then we can measure the orientation of the ambient field from the angle of bilateral symmetry. Based on the polarization observations, we assume the northern SNR is at a further distance. We define the ambient magnetic field at this distance to have a different orientation than the one for the southern, foreground SNR (see the green lines in Figure~3).


\section{Conclusions}
We find a good match between data and model for a distance to the southern SNR of 491 pc (radius=19 pc) and a transition distance of 487~pc. The published distances to the Cygnus Loop have a range 0.46--0.64 kpc \citep[and references therein]{2012AdSpR..49.1313F}.

Note that our model does not include a Faraday screen or age differences between the northern and southern SNRs and thus, our model does not show the depolarization found in the GALFACTS data. For these distances, we find that there would be interaction between the two SNRs.

\begin{table}
\center
\begin{tabular}{|c|c|c|c|c|c|}
\hline 
 & $\alpha$ & $\delta$ & $l$ ($^{\circ}$) & $b$ ($^{\circ}$) & size ($^{\circ}$)\tabularnewline
\hline 
\hline 
Northern SNR & 20h51.6m & 31$^{\circ}$3.0' & 74.3 & -8.4 & 1.40\tabularnewline
\hline 
Southern SNR & 20h48.8m & 29$^{\circ}$47.3' & 73.0 & -8.7 & 1.07\tabularnewline
\hline 
\end{tabular}
\caption{Regions used in the analysis and modelling.}
\end{table}

\begin{figure}[!tbp]
  \centering
  \begin{minipage}[b]{0.5\textwidth}
    \includegraphics[width=\textwidth]{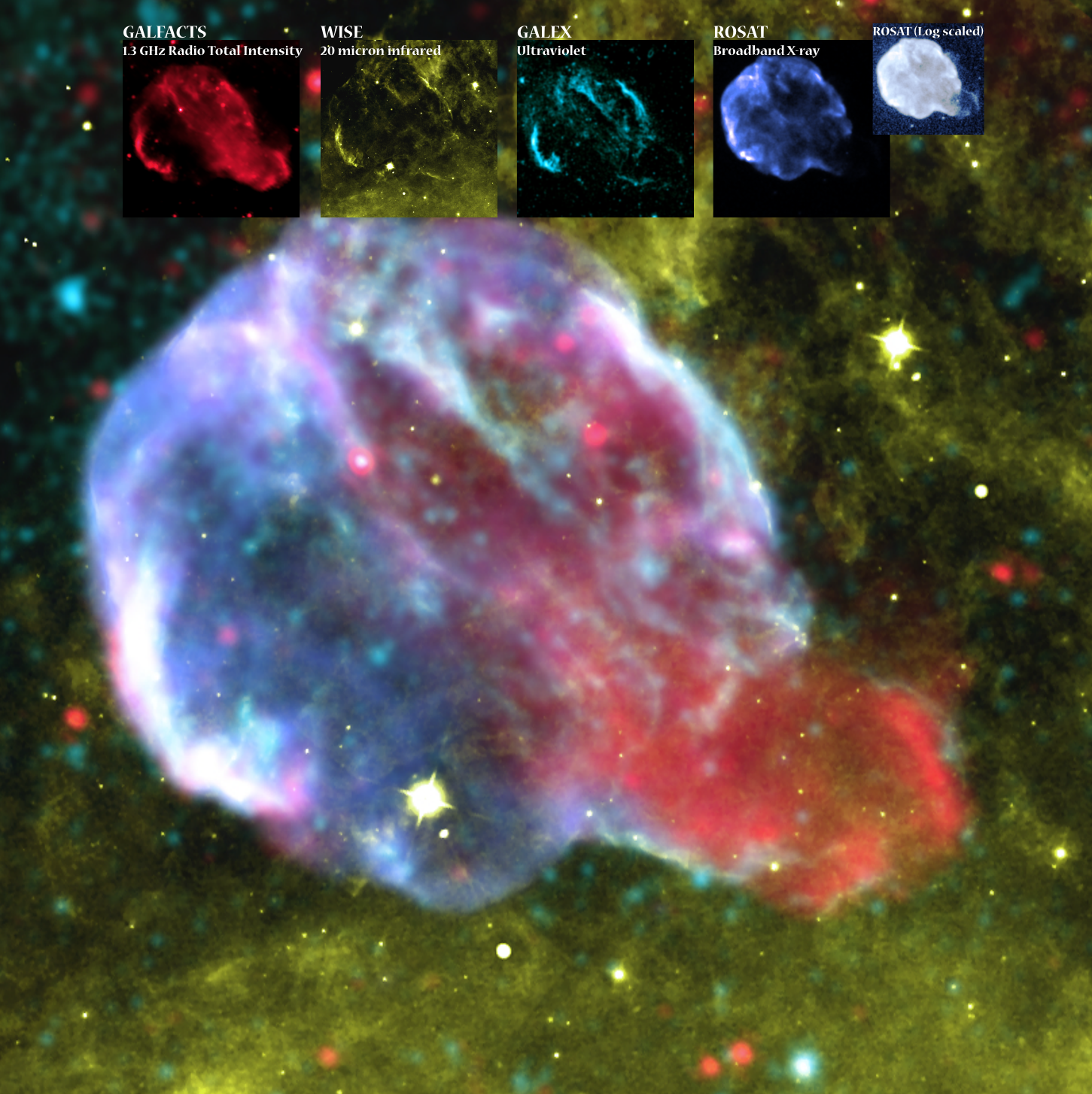}
    \caption{Colour image made by combining GALFACTS (red), WISE (green), GALEX (cyan), and ROSAT (blue) data. Individual images are shown above. Note that the image
is displayed in Galactic coordinates.}
  \end{minipage}
  \hfill
  \begin{minipage}[b]{0.4\textwidth}
    \includegraphics[width=\textwidth]{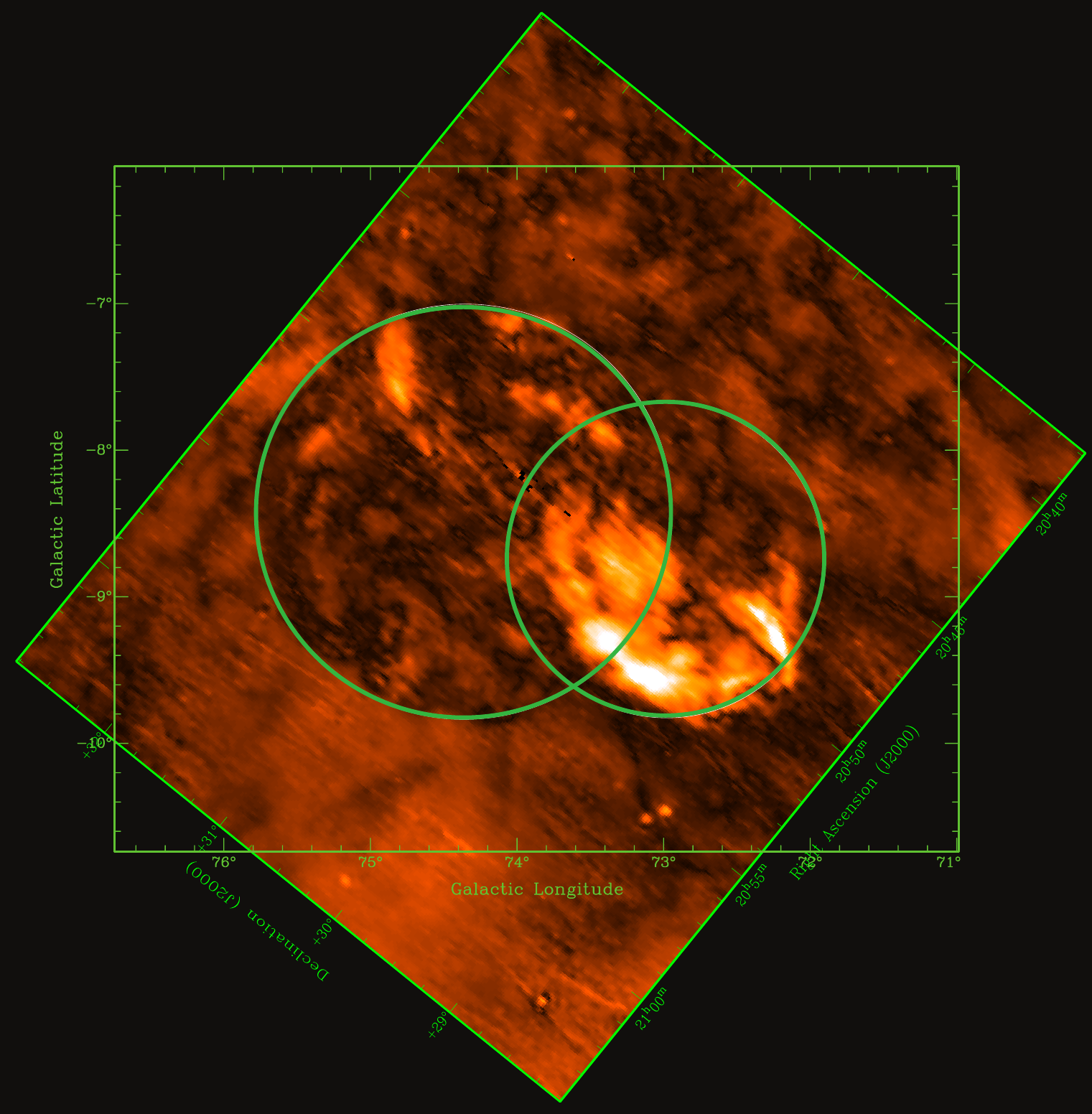}
    \caption{Data shown in comparison to the best fit 2-SNR model. The green lines on the Model Total Intensity image indicate the orientations of the ambient magnetic fields for the background (at the distance of the northern SNR) and foreground (at the distance of the southern SNR).}
  \end{minipage}
\end{figure}



\begin{figure}[!tbp]
  \centering
  \begin{minipage}[b]{0.5\textwidth}
    \includegraphics[width=\textwidth]{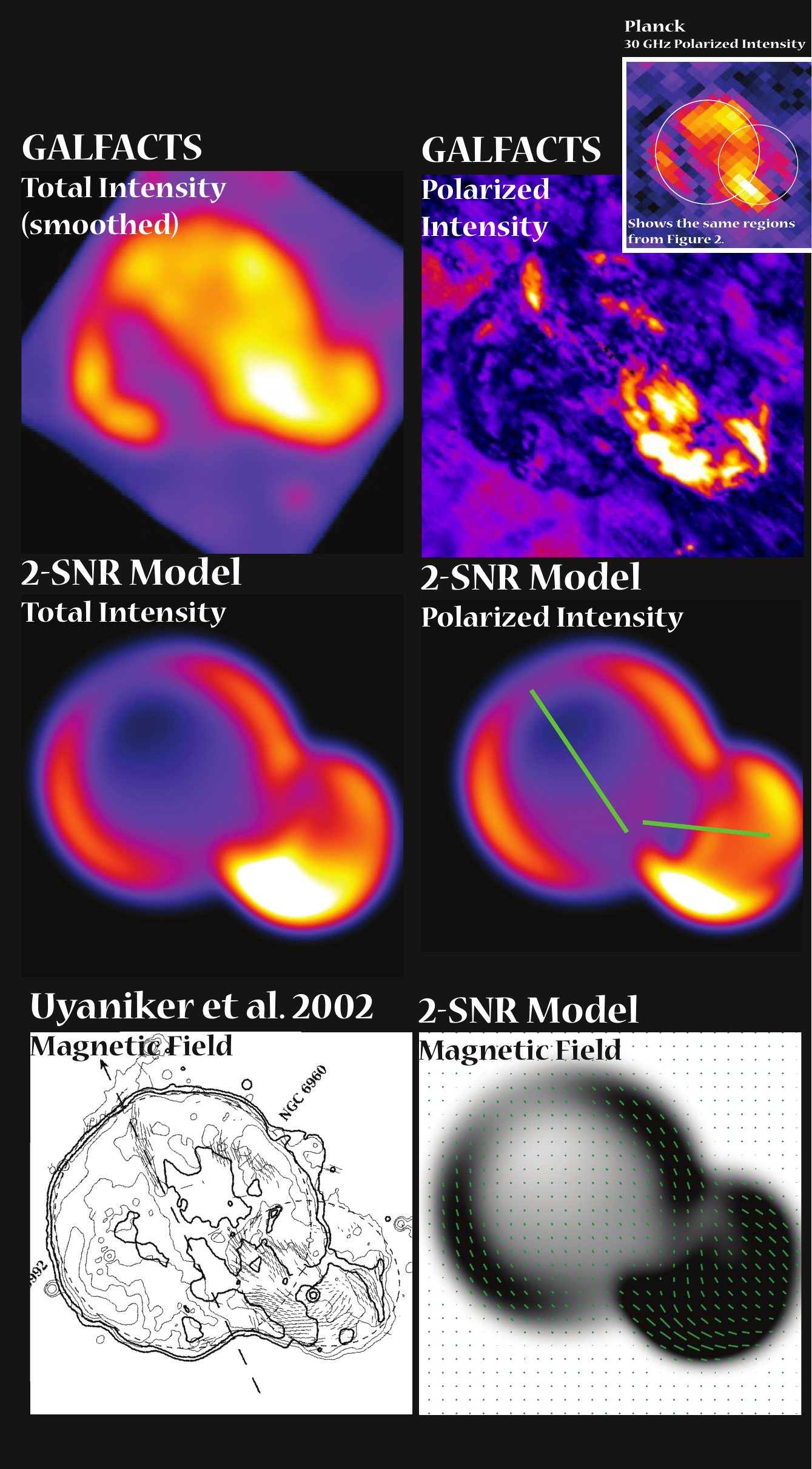}
    \caption{Data shown in comparison to the best fit 2-SNR model. The green lines on the Model Total Intensity image indicate the orientations of the ambient magnetic fields for the background (at the distance of the northern SNR) and foreground (at the distance of the southern SNR).}
  \end{minipage}
  \hfill
  \begin{minipage}[b]{0.4\textwidth}
    \includegraphics[width=\textwidth]{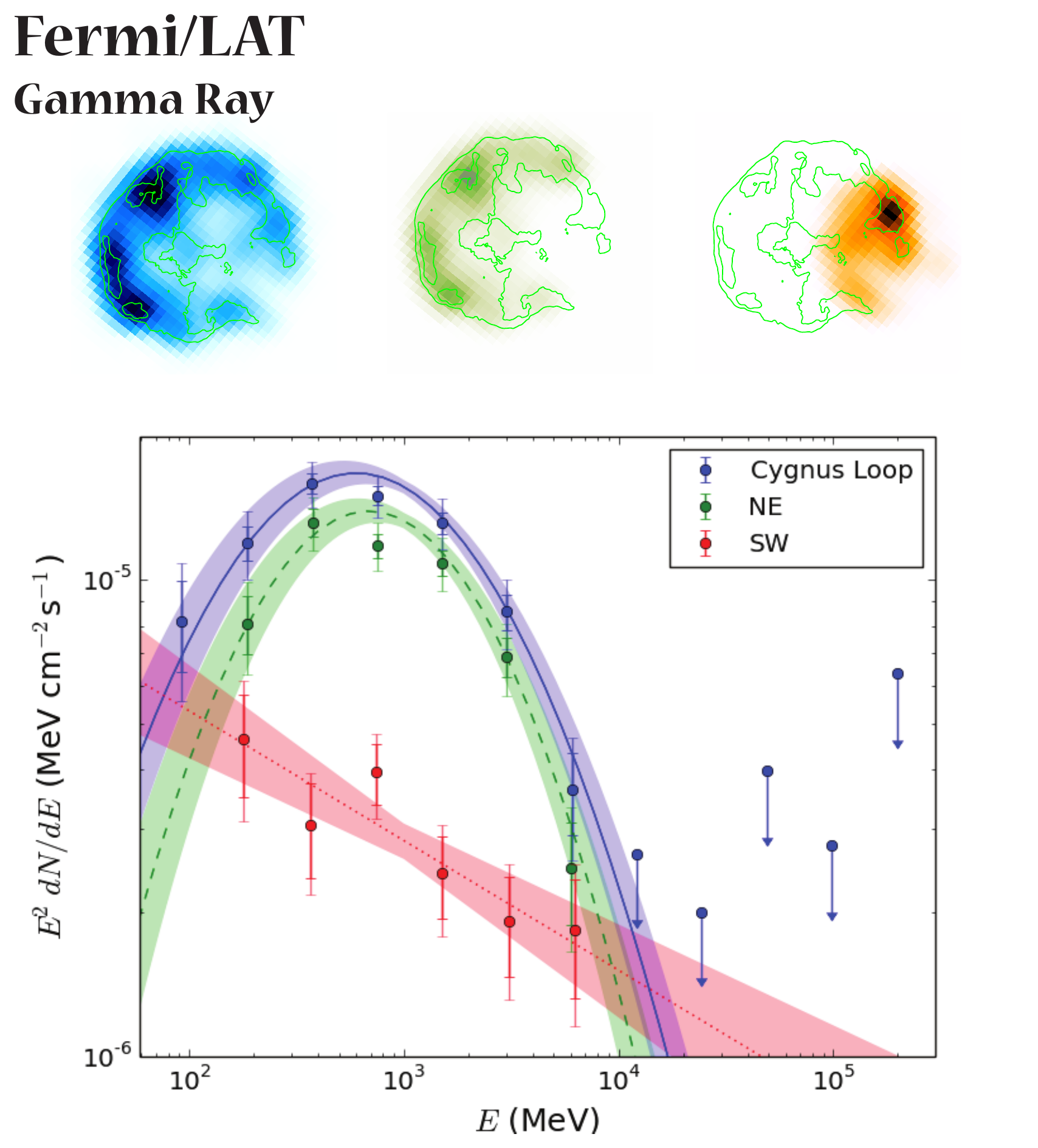}
    \caption{The images above show the image templates used to derive spectra from the Fermi/LAT data. The entire Cygnus Loop is shown (blue); using the regions defined by the GALFACTS data, the northern (NE) SNR is isolated and shown in green and the southern (SW) SNR is isolated and shown in red. The energy spectra from these regions is shown below.  The statistical uncertainty range of the best model is shown as a shaded area.  Spectral points include statistical uncertainties (solid bars) and systematic uncertainties (shaded bars).}
  \end{minipage}
\end{figure}



\small  
%
\section*{Acknowledgments}   
%
This research was supported by the Natural Sciences \&
Engineering Research Council of Canada,
the Canada Foundation for Innovation and
the Manitoba Research and Innovation Fund. \\

\bibliographystyle{aj}
\small
\bibliography{proceedings}
\end{document}